\documentclass[onecolum, notitlepage, showpacs,amsmath,amssymb,superscriptaddress]{revtex4-2}

\usepackage{physics}
\usepackage{graphicx} 
\usepackage[caption=false, labelformat=simple]{subfig}
\usepackage{dcolumn}
\usepackage{csquotes}
\usepackage[colorlinks=true,linkcolor = {blue}, citecolor = {blue}, urlcolor = {blue}]{hyperref}
\usepackage{xcolor}

\usepackage[export]{adjustbox}

\begin{document}

\title{Accurate simulation of Efimov physics in ultracold atomic gases with realistic three-body multichannel interactions}

\author{J. van de Kraats}
\email[Corresponding author:  ]{j.v.d.kraats@tue.nl}
\affiliation{Eindhoven University of Technology, P. O. Box 513, 5600 MB Eindhoven, The Netherlands}
\author{S.J.J.M.F. Kokkelmans}
\affiliation{Eindhoven University of Technology, P. O. Box 513, 5600 MB Eindhoven, The Netherlands}
\date{\today}

\begin{abstract}

We give a detailed and self-contained description of a recently developed theoretical and numerical method for the simulation of three identical bosonic alkali-metal atoms near a Feshbach resonance, where the Efimov effect is induced. The method is based on a direct construction of the off-shell two-body transition matrix from exact eigenfunctions of the embedded two-body Hamiltonians, obtained using realistic parameterizations of the interaction potentials which accurately reproduce the molecular energy levels. The transition matrix is then inserted into the appropriate three-body integral equations, which may be efficiently solved on a computer. We focus especially on the power of our method in including rigorously the effects of multichannel physics on the three-body problem, which are usually accounted for only by various approximations. We demonstrate the method for \textsuperscript{7}Li, where we recently showed that a correct inclusion of this multichannel physics resolves the long-standing disagreement between theory and experiment regarding the Efimovian three-body parameter. We analyze the Efimovian enhancement of the three-body recombination rate on both sides of the Feshbach resonance, revealing strong sensitivity to the spin structure of the model thus indicating the prevalence of three-body spin-exchange physics. Finally, we discuss an extension of our methodology to the calculation of three-body bound-state energies.

\end{abstract}

\maketitle

\section{Introduction}
\label{sec:intro}

The Efimov effect constitutes a central result of few-body quantum physics, describing the sudden appearance of an infinite tower of three-body bound states in any system with short-range but resonant pairwise interactions \cite{Efimov1970, Efimov1971, Braaten2006, Naidon2017, Greene2017, Incao2018}. In the simplest case of zero-range contact interactions, parameterized by an $s$-wave scattering length $a$, the associated spectrum of Efimov states has discrete scale invariance relative to a characteristic scaling factor $\lambda = e^{\pi/s_0}$, where $s_0 = 1.00624$ is a universal dimensionless number. In configuration space this invariance originates from an induced attractive interaction scaling precisely as $1/R^2$, where $R$ is the size of the three-body system. For more realistic interactions with a finite range, the scale invariance of the Efimov spectrum is broken through the appearance of a characteristic length scale, now typically referred to as the three-body parameter. Initially, the three-body parameter was expected to be highly sensitive to short-range details of the system, thus making any reasonably universal prediction for its value from two-body physics unlikely unless the interaction could be characterized by a large universal effective range \cite{Incao2009, Petrov2004}.

Interest in the Efimov effect found a resurgence once it was realized that it could be experimentally accessed in trapped ultracold atomic gases, where precision control of the scattering length is enabled through the use of magnetic Feshbach resonances \cite{Feshbach1958, Feshbach1962, Chin2010}. In this scheme, an Efimov state can be experimentally characterized by scanning the scattering length on both the repulsive and attractive side of the resonance, whilst monitoring the loss of atoms from the trap. Then, Efimovian enhancement of three-body recombination processes leads to a log-periodic modulation of the atomic loss rate as a function of scattering length, at a characteristic frequency set by the Efimov constant $s_0$  \cite{Esry1999, Braaten2006, Naidon2017}. The three-body parameter, typically expressed in terms of  the characteristic scattering lengths $a_{\pm}$ and trimer widths $\eta_{\pm}$, can then be directly extracted from the absolute phase of the oscillations.

By exploiting Feshbach resonances, the three-body parameter has been measured in a wide array of different atomic species amicable to laser cooling. Surprisingly, these experiments have revealed that the three-body parameter for broad Feshbach resonances is universally determined as $a_- \approx -9.7 \ r_{\mathrm{vdW}}$, where $r_{\mathrm{vdW}}$ is the range of the two-body interaction \cite{Kraemer2006, Berninger2011, Wild2012}. This van der Waals universality contradicts the earlier expectation that the three-body parameter would be sensitive also to the nonuniversal atomic structure, thus reigniting theoretical interest in the Efimov effect. Detailed studies of the three-body problem subsequently revealed that van der Waals universality originates from a characteristic repulsive barrier in the three-body potential, which shields the particles from probing the non-universal short range \cite{Wang2012, Naidon2014}. Later, this framework was extended further to show how the three-body parameter obeys distinct universality classes dependent on the shape of the two-body interaction \cite{Naidon2014_2}.

The studies of van der Waals universality provide a satisfactory picture of the Efimov effect for broad Feshbach resonances, where the effective range of the two-body interaction is small \cite{Chin2010}. In the opposite limit of a narrow Feshbach resonance, with large effective range, the three-body parameter is also expected to be universally determined \cite{Petrov2004, Gogolin2008}, although experimental verification in this regime is lacking due to difficulties in working with very small magnetic field ranges. Experimental data is available however, in the much more challenging regime of intermediate resonance widths, where no single dominant two-body length scale can be defined. Here, theoretical models based on multichannel two-body interactions generally predict an increase of the value of $\abs*{a_-}$ as the resonance narrows, which then smoothly connects to the analytic regime of very narrow resonances where $a_-$ starts to scale directly with the effective range \cite{Schmidt2012, Langmack2018, Secker2021_2C, Kraats2023}. Experimental measurements of $a_-$ obtained near intermediate Feshbach resonances in \textsuperscript{39}K are generally in line with this prediction \cite{Roy2014, Chapurin2019, Etrych2023}, giving credence to the fundamental assumptions about the relevant microscopic physics on which these models are based. Puzzling however, has been the case of \textsuperscript{7}Li, where the value of $\abs{a_-}$ has been measured to be \textit{smaller} than the universal van der Waals value, even though the associated Feshbach resonances are actually narrower than those used in potassium \cite{Gross2009, Gross2010, Gross2011, Dyke2013}. Furthermore, most known theoretical models, even those that attempted to include non-universal details of the two-body interaction, could not reproduce these findings to any satisfactory degree \cite{Yudkin2021, Incao2022, Yudkin2023_abc}. This \textit{lithium few-body puzzle}, became additionally interesting when new measurements revealed the existence of an Efimov state above the atom-dimer scattering threshold, which was subsequently found to originate from a universal reshape of the three-body interaction for intermediate to narrow Feshbach resonances \cite{Yudkin2024}. At roughly the same time, the anomalous value of $a_-$ was finally reproduced in a numerical calculation, which included the effects of strong three-body spin-exchange interactions, thus resolving the long-standing discrepancy between theory and experiment \cite{Kraats2024}.

Following up on the developments above, in this article we discuss in detail a theoretical and numerical method for the three-body problem which allows for the inclusion of all spin degrees of freedom of the interacting atoms, as was required in resolving the lithium three-body puzzle regarding the three-body parameter \cite{Kraats2024}. We subsequently discuss the extension of this method to also calculate the energy of Efimov resonances, which are finding increased experimental relevance \cite{Yudkin2024}, and apply the method for a study of three-body recombination in \textsuperscript{7}Li on both sides of the Feshbach resonance, including also the regime of positive scattering length where a shallow dimer state exists. The paper will be structured as follows. In Sec. \ref{sec:theory} we discuss in detail our theoretical and numerical methods for obtaining both the three-body recombination rates and the energies of Efimov states in systems of three interacting ultracold alkali-metal atoms. Then in Sec. \ref{sec:results} we give and discuss the results we obtain upon applying this model to the \textsuperscript{7}Li system specifically. Finally we conclude this paper in Sec. \ref{sec:conc}, and give some outlook for further research.

\section{Method}
\label{sec:theory}

In this section we describe in detail our multichannel method for calculating both the three-body recombination rate and three-body bound state energy near a Feshbach resonance, for three bosonic alkali-metal atoms with mass $m$ labelled with Latin indices $j = 1,2,3$. We work in Jacobi coordinates \cite{Sitenko1991}, meaning that the three-body system is subdivided into a pair + a third particle. We specify the choice of pair by a Greek index $\alpha$, equal to the index $j$ of the third particle. For example if $\alpha = 3$, and particles have individual positions $\vb{x}_j$ relative to some arbitrary origin, then the set of Jacobi coordinates is defined as,
\begin{align}
\begin{split}
\vb{r}_3 &= \vb{x}_1 - \vb{x}_2, \\
\vb*{\rho}_3 &= \vb{x}_3 - \frac{1}{2} (\vb{x}_1 + \vb{x}_2),\\
\vb{R} &= \frac{1}{3} (\vb{x}_1 + \vb{x}_2  +  \vb{x}_3 ).
\end{split}
\end{align}
Similarly, if particles have individual momenta $\tilde{\vb{k}_j}$,
\begin{align}
\begin{split}
\vb{k}_3 &= \frac{1}{2} \left(\tilde{\vb{k}}_1 -  \tilde{\vb{k}}_2 \right), \\
\vb{p}_3 &= \frac{2}{3} \tilde{\vb{k}}_3 - \frac{1}{3} (\tilde{\vb{k}}_1 + \tilde{\vb{k}}_2), \\
\vb{K} &= \tilde{\vb{k}}_1 +\tilde{\vb{k}}_2 +\tilde{\vb{k}}_3.
\end{split}
\end{align}
We refer to $\vb{r}_3, \vb{k}_3$ as the \textit{dimer} separation/momentum and $\vb*{\rho}_3, \vb{p}_3$ as the \textit{atom-dimer} separation/momentum. We assume translational invariance of the total three-body system, meaning that the trapping potential is essentially flat in the regime of interest, such that it is valid to set $\vb{R} = 0$ or $\vb{K} = 0$. We also note that the particles are indistinguishable, so any choice of Jacobi set is arbitrary, and we will see later that all equations are by nature symmetrized with respect to different choices of $\alpha$.

\subsection{Hamiltonian}

\begin{figure}[t]
\includegraphics{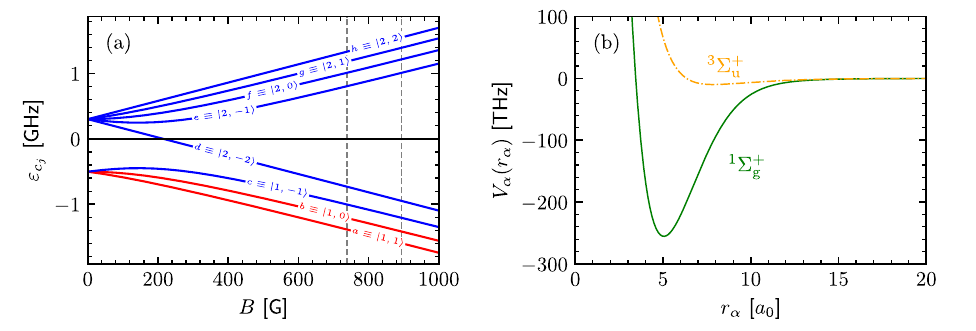}
\caption{\label{fig:hf_pot} Characteristics of the coupled-channels few-body problem \cite{Stoof1988}. In (a) we show the spectrum $\varepsilon_{c_j}$ of channel states for a single \textsuperscript{7}Li atom, using the labelling convention as explained in the main text. States highlighted in red correspond to the initial states studied in this work, which possess Feshbach resonances at the magnetic field values indicated by the vertical dashed lines. In (b) we plot the two-body singlet and triplet potential energy surfaces $V_{\alpha}^{S}(r_{\alpha})$, as formulated for \textsuperscript{7}Li in Ref.~\cite{Julienne2014}.}
\end{figure}

In absence of interactions, a system of three alkali-metal atoms moving in an external magnetic field $\vb{B}$ is described by the sum of single-particle Hamiltonians \cite{Secker2021_2C, Li2022},
\begin{align}
\begin{split}
\hat{H}_0 =  \sum_{j=1}^3 \left[\hat{T}_j + \hat{H}_j^{\mathrm{hf}}  + \hat{H}_j^{\mathrm{Z}} (\vb{B})\right].
\end{split}
\label{eq:H}
\end{align}
Here $\hat{T}_j$ is the kinetic energy operator for particle $j$, and $\hat{H}_j^{\mathrm{hf}}, \hat{H}_j^{\mathrm{Z}} (\vb{B})$ are the single-particle hyperfine and Zeeman Hamiltonians which account for the spin state of the particle. They are defined explicitly as,
\begin{align}
\begin{split}
\hat{H}_j^{\mathrm{hf}} = A_j^{\mathrm{hf}} \hat{\vb{s}}_j \cdot \hat{\vb{i}}_j, \qquad
\hat{H}_j^{\mathrm{Z}}(\vb{B}) = \left(\gamma_j^n \hat{\vb{i}}_j + \gamma_j^e \hat{\vb{s}}_j \right) \cdot \vb{B},
\end{split}
\end{align}
where $\hat{\vb{s}}_j$ and $\hat{\vb{i}}_j$ are the electronic and nuclear spin operators, $A_j^{\mathrm{hf}}$ is the hyperfine energy, and $\gamma_j^{\mathrm{n/e}}$ are the nuclear and electronic gyromagnetic ratios. Upon diagonalizing $\hat{H}_j^{\mathrm{hf}} + \hat{H}_j^{\mathrm{Z}}$ one obtains a spectrum of single particle eigenstates with magnetic field dependent energy $\varepsilon_{c_j}$, as shown in Fig.~\ref{fig:hf_pot}(a) for the case of \textsuperscript{7}Li. By convention, these states are labelled in alphabetical order with respect to increasing energy, or alternatively by the hyperfine state $\ket*{f_j, m_{f_j}}$ to which they adiabatically connect when $\abs{\vb{B}} \rightarrow 0$. Here $\hat{\vb{f}}_j = \hat{\vb{s}}_j + \hat{\vb{i}}_j$, and $m_{f_j}$ is the associated projection on the magnetic field direction. 

The total Hamiltonian including interactions is written as,
\begin{align}
\begin{split}
\hat{H} = \hat{H}_0 + \sum_{\alpha} \hat{V}_{\alpha},
\end{split}
\end{align}
where $\hat{V}_{\alpha}$ represents the two-body interaction for the pair $\alpha$. For neutral atoms at large interparticle separations ($r_{\alpha} \rightarrow \infty$), the dominant contribution to the interaction is given by the attractive van der Waals interaction, i.e. $V_{\alpha}(r_{\alpha}) \rightarrow - C_6/r_{\alpha}^6$. Here the dispersive coefficient $C_6$ defines the characteristic van der Waals length scale $r_{\mathrm{vdW}} = \frac{1}{2} \left(\frac{m C_6}{\hbar^2} \right)^{\frac{1}{4}} \approx 32.4863 a_0$. At shorter separations, where $r_{\alpha} \ll r_{\mathrm{vdW}}$, the overlapping wave functions of the valence electrons are projected into either singlet ($S = 0$) or triplet ($S = 1$) spin configurations. In this regime, the interaction accordingly decomposes into two components,
\begin{align}
\begin{split}
\hat{V}_{\alpha} = \hat{V}_{\alpha}^{0} \hat{\mathcal{P}}_{\alpha}^0 + \hat{V}_{\alpha}^1 \hat{\mathcal{P}}_{\alpha}^1,
\end{split}
\end{align}
where $\hat{\mathcal{P}}_{\alpha}^{S}$ projects the pair into the electronic spin state $S$. The detailed shape of the singlet and triplet molecular potentials $\hat{V_{\alpha}}^{S}$ is generally very complicated, depending on non-universal detail of the atomic structure. We use a realistic parameterization of the molecular potential energy surface obtained by fitting two-body coupled-channels calculations to low-energy scattering and spectroscopic data \cite{Julienne2014}, as plotted in Fig.~\ref{fig:hf_pot}(b).

\subsection{Three-body basis}
\label{sec:basis}

In a uniform system, the state of the paired particles can be represented by the plane wave state $\ket*{\vb{k}_{\alpha}}$ \cite{Taylor2006}, where,
\begin{equation}
\braket*{\vb{r}_{\alpha}}{\vb{k}_{\alpha}} = \frac{1}{(2 \pi)^{\frac{3}{2}}} e^{i \vb{k}_{\alpha} \cdot \vb{r}_{\alpha}}, \qquad \int d^3\vb{k}_{\alpha} \dyad*{\vb{k}_{\alpha}}{\vb{k}_{\alpha}} = \mathbb{I}
\end{equation}
These basis states are not very efficient for the present computations since they do not exploit the spherical symmetry of the pairwise interaction. Instead, we work in a partial wave basis $\ket*{k_{\alpha}, l_{\alpha}, m_{l_{\alpha}}}$, where $l_{\alpha}$ is the orbital angular momentum quantum number of the dimer subsystem, and $m_{l_{\alpha}}$ the associated projection on the quantization axis. The normalization reads \cite{Glockle1983},
\begin{align}
\begin{split}
&\braket*{k_{\alpha}', l_{\alpha}', m_{l_{\alpha}}'}{k_{\alpha}, l_{\alpha}, m_{l_{\alpha}}} = \frac{1}{4 \pi} \frac{\delta(k_{\alpha}' -k_{\alpha})}{k_{\alpha}' k_{\alpha}} \delta_{l_{\alpha} l_{\alpha}'} \delta_{m_{l_{\alpha}} m_{l_{\alpha}}'}, \\  &\sum_{l_{\alpha}, m_{l_{\alpha}}}\int_0^{\infty} dk \  4\pi k_{\alpha}^2 \ket{k_{\alpha}, l_{\alpha}, m_{l_{\alpha}}} \bra{k_{\alpha}, l_{\alpha}, m_{l_{\alpha}}} =\mathbb{I}.
\end{split}
\label{eq:basisnorm}
\end{align}
As follows from the Wigner-Eckart theorem, any spherically symmetric operator in this basis is independent of $m_{l_{\alpha}}$ and diagonal in $l_{\alpha}$. The position and momentum projections are,
\begin{align}
\begin{split}
\braket*{\vb{r}_{\alpha}}{k_{\alpha}, l_{\alpha}, m_{l_{\alpha}}} &= \frac{i^{l_{\alpha}}}{\sqrt{2} \pi} \frac{\hat{j}_{l_{\alpha}}(k_{\alpha}r_{\alpha})}{k_{\alpha}r_{\alpha}} Y_{l_{\alpha}}^{m_{l_{\alpha}}}(\vu{r}_{\alpha}), \\\braket*{\vb{k}_{\alpha}'}{k_{\alpha}, l_{\alpha}, m_{l_{\alpha}}} &= \frac{1}{2 \sqrt{\pi}}  \frac{\delta(k_{\alpha}-k_{\alpha}')}{k_{\alpha} k_{\alpha}'} Y_{l_{\alpha}}^{m_{l_{\alpha}}}(\vu{k}_{\alpha}'),
\end{split}
\label{eq:PartWaveToPosAndMom}
\end{align}
where $\hat{j}_{l_{\alpha}}(k_{\alpha}r_{\alpha})$ is the Riccati-Bessel function \cite{Taylor2006}, and $Y_{l_{\alpha}}^{m_{l_{\alpha}}}(\vu{k}_{\alpha})$ a spherical harmonic.

The extension to the three-body sector follows naturally by defining an equivalent basis state $\ket*{p_{\alpha}, \lambda_{\alpha}, m_{\lambda_{\alpha}}}$ for the atom-dimer degree of freedom. As before, $p_{\alpha}$ is the atom-dimer momentum, $\lambda_{\alpha}$ the atom-dimer angular momentum, and $m_{\lambda_{\alpha}}$ its projection. Total rotational symmetry of the system then motivates coupling the dimer and atom-dimer angular momenta to obtain the total angular momentum $\vb{L} = \vb{l}_{\alpha} + \vb*{\lambda}_{\alpha}$ \cite{Glockle1983}. The associated basis states are formed as,
\begin{align}
\begin{split}
\ket*{l_{\alpha},\lambda_{\alpha}, L, M_{L}} = \sum_{m_{l_{\alpha} } m_{\lambda_{\alpha} }} \braket*{l_{\alpha} , m_{l_{\alpha} }, \lambda_{\alpha} , m_{\lambda_{\alpha} }}{L,M_{L}} \ket*{l_{\alpha} , m_{l_{\alpha} }, \lambda_{\alpha} , m_{\lambda_{\alpha} }}.
\end{split}
\label{eq:PartWaveBaseRecouple}
\end{align}
Here $\braket*{l_{\alpha} , m_{l_{\alpha} }, \lambda_{\alpha} , m_{\lambda_{\alpha} }}{L,M_{L}}$ is a Clebsch-Gordan coefficient. Spherical symmetry dictates that $L, M_L$ are both good quantum numbers. Including also the absolute momenta, we now write the total three-body partial wave basis as $\ket*{k_{\alpha}p_{\alpha}(l_{\alpha} \lambda_{\alpha}) L M_L}$ \cite{Glockle1983}. It is an eigenstate of the kinetic energy operator,
\begin{align}
\begin{split}
\sum_j T_j \ket*{k_{\alpha}p_{\alpha}(l_{\alpha} \lambda_{\alpha}) L M_L} = \left(\frac{\hbar^2 k_{\alpha}^2}{m} + \frac{3}{4} \frac{\hbar^2 p_{\alpha}^2}{m} \right) \ket*{k_{\alpha}p_{\alpha}(l_{\alpha} \lambda_{\alpha}) L M_L}.
\end{split}
\end{align}
Going forward we assume $L = M_L = 0$, as this term gives the dominant contribution to the three-body recombination rate at low energies \cite{Esry2001}. This immediately enforces that $\lambda_{\alpha} = l_{\alpha}$, so we only need a single orbital quantum number $l_{\alpha}$ to specify the three-body state.

To describe the spin state of particles, we do not work directly with the hyperfine states $\ket*{f_j, m_{f_j}}$ as defined in the previous section, but instead use the basis of atomic or Zeeman single-particle states $\ket*{\xi_j} = \ket*{m_{s_j}, m_{i_j}}$, where $m_{s_j}, m_{i_j}$ are the magnetic projections of the electronic and nuclear spin respectively. In the three-body sector, the only rigorously conserved quantum number is the total magnetic spin projection $M_F$,
\begin{align}
M_F = \sum_{j = 1}^3 m_{s_j} + \sum_{j = 1}^3 m_{i_j} \equiv \sum_{j = 1}^3 m_{f_j}.
\label{eq:BasisCond}
\end{align}

\begin{figure}[t]
\includegraphics{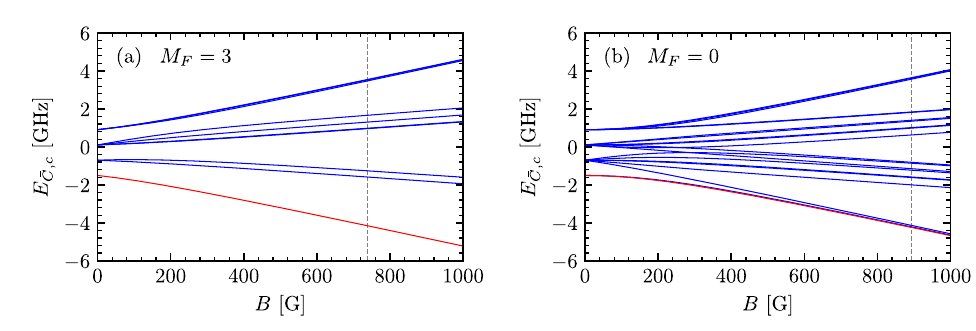}
\caption{\label{fig:3BodyThresholds} Three-body channel energies $E_{\bar{C}, c}$ for \textsuperscript{7}Li as a function of magnetic field, for two values of the total magnetic quantum number $M_F$ relevant to this work. The state with lowest energy, setting the three-body scattering threshold, is drawn in red. Vertical dashed lines show the location of the Feshbach resonance studied in this work.}
\end{figure}

This means that, for any particular choice of Jacobi set $\alpha$, the spin projection for the dimer particles equals $M_F - m_{f_{\alpha}}$, which is known to be preserved by $V_{\alpha}$. Hence the spin state of the third particle, $\ket*{\xi_{\alpha}}$, will determine the set of coupled spin states for the dimer subsystem, henceforth notated as $\ket*{\Xi_{\alpha}}$. For identical particles, the dimension of this set can be reduced by symmetrization,
\begin{align}
\begin{split}
\ket*{\bar{\Xi}_{\alpha}} = \frac{1 + (-1)^l \hat{P}_{\alpha}}{\sqrt{2\left(1 + \delta_{\alpha} \right)}} \ket*{\Xi_{\alpha}},
\end{split}
\end{align}
Here $\hat{P}_{\alpha}$ denotes a permutation operator that exchanges the two paired particles, and $\delta_{\alpha}$ equals unity when the two particles are in identical states and zero otherwise. The factor $(-1)^l$ corrects for the spatial parity due to the angular momentum. Now, in the three-body sector, we form the matrix $\sum_{j=1}^3 \matrixel*{\bar{\Xi}_{\alpha}', \xi_{\alpha}'}{\hat{H}_j^{\mathrm{hf}} + \hat{H}_j^{\mathrm{Z}}}{\bar{\Xi}_{\alpha}, \xi_{\alpha}}$, which is diagonalized to obtain eigenstates $\ket*{\bar{C}_{\alpha}, c_{\alpha}}$, referred to as channel states, with energies $E_{\bar{C}_{\alpha}, c_{\alpha}} = \varepsilon_{\bar{C}_{\alpha}} + \varepsilon_{c_{\alpha}}$. In Fig.~\ref{fig:3BodyThresholds} we illustrate this three-body channel spectrum for \textsuperscript{7}Li, given two distinct values of the total projection $M_F$. Here one observes clearly how $M_{F}$ sets the total number of coupled spin states, and accordingly influences the complexity of the three-body calculation. For the remainder of this section, we will set $E = 0$ at the three-body scattering threshold, given by the lowest three-body channel energy as highlighted in red in Fig.~\ref{fig:3BodyThresholds}.

This concludes the definition of the three-body basis, $\ket*{k_{\alpha}p_{\alpha}l_{\alpha};\bar{C}_{\alpha} c_{\alpha}} \equiv \ket*{k_{\alpha}p_{\alpha}(l_{\alpha} l_{\alpha}) 0 0} \otimes \ket*{\bar{C}_{\alpha} c_{\alpha}}$. For notational brevity, we will suppress the Jacobi index $\alpha$ on momenta and quantum numbers going forward, as the appropriate Jacobi index can always be inferred from the surrounding equation.

\subsection{Off-shell two-body transition matrix}

As input to the three-body calculation, we require knowledge of the two-body transition operator for a given partial wave $l$ and third particle state $c$, denoted as $\hat{t}_{\alpha}^{l c}(z)$. Here $z$ is the two-body energy, which will have to be evaluated \textit{off-shell} as the two-body subsystem may lose or gain energy by interacting with the third particle \cite{Sitenko1991}. In general, the transition operator can be defined through the Lippmann-Schwinger equation, which has the formal solution \cite{Taylor2006},
\begin{align}
\hat{t}_{\alpha}^{l c}(z) = \hat{V}_{\alpha} + \hat{V}_{\alpha} \hat{g}_{\alpha}^{l c}(z) \hat{V}_{\alpha},
\label{eq:Tmatdef}
\end{align}
with $\hat{g}_{\alpha}^{l c}(z) = (z - \hat{h}_{\alpha}^{lc})^{-1}$ the two-body Green's operator, and two-body Hamiltonian $\hat{h}_{\alpha}^{l c}$. Defining two-body eigenstates as $\hat{h}_{\alpha}^{lc}\ket*{\psi_{l c}^{\nu}}  = \varepsilon_{lc}^{\nu} \ket*{\psi_{l c}^{\nu}} $, with state index $\nu$, the Green's operator can be expanded as,
\begin{align}
\begin{split}
\hat{g}_{\alpha}^{l c}(z) = \sum_{\nu} \frac{\ket*{\psi_{l c}^{\nu}} \bra*{\psi_{l c}^{\nu}}}{z - \varepsilon_{lc}^{\nu}},
\end{split}
\end{align}
Inserting this expansion into Eq. \eqref{eq:Tmatdef} and evaluating in the two-body partial wave and spin basis gives,
\begin{align}
\matrixel*{k'l; \bar{C}'}{\hat{t}_{\alpha}^{l c}(z)}{k l; \bar{C}} &= \matrixel*{k'l ; \bar{C}'}{\hat{V}_{\alpha}}{kl ; \bar{C}} + \sum_{\nu} \matrixel*{k'l ; \bar{C}'}{\hat{V}_{\alpha}\frac{\ket*{\psi_{l c}^{{\nu}}} \bra*{\psi_{l c}^{{\nu}}}}{z - \varepsilon_{lc}^{\nu}}\hat{V}_{\alpha}}{kl ; \bar{C}}.
\label{eq:LSbasis}
\end{align}
Here we have integrated away the angular degrees of freedom, such that the inconsequential quantum number $m_l$ drops out. We will now use Eq.~\eqref{eq:LSbasis} to construct the transition matrix from the two-body eigenfunctions $\ket*{\psi_{l c}^{{\nu}}}$, which we obtain by solving the associated quantum mechanical scattering problem in position space. First, we exploit spherical symmetry of $\hat{V}_{\alpha}$ by expanding the wave function as,
\begin{align}
\braket*{r}{\psi_{l c}^{{\nu}}} = \sum_{\bar{C}} \frac{u_{lc\bar{C}}^{\nu}(r)}{r}  \ket*{\bar{C}},
\end{align}
where $u_{lc\bar{C}}^{\nu}(r)$ is the radial wave function for each individual two-body scattering channel. It obeys the radial Schr\"odinger equation,
\begin{align}
\begin{split}
\left[\frac{\hbar^2}{m} \left( - \dv[2]{r} + \frac{l(l+1)}{r^2} \right) + V_{\alpha}(r)\right] u_{lc\bar{C}}^{\nu}(r) = \left(\varepsilon_{lc}^{\nu} - \varepsilon_{\bar{C}} \right) u_{lc\bar{C}}^{\nu}(r).
\end{split}
\label{eq:SchrodRad}
\end{align}
We solve Eq.~\eqref{eq:SchrodRad} via a discrete variable representation (DVR) \cite{Light1985}, with a transformed grid in $r$ for increased efficiency \cite{Fattal1996, Willner2004}. Although such methods are typically used to calculate molecular binding energies, constructing the transition matrix requires not just the bound states but also the scattering states, for which $\varepsilon_{lc}^{\nu} - \varepsilon_{\bar{C}} > 0$. In the DVR these states are not exact, due to the discretization of the coordinate space via a hard-box boundary condition at some (large) two-body separation $r_b$, which should far exceed all other length scales. For a detailed study of the effects of this box on the resulting transition matrix and three-body observables, we refer to ref. \cite{Secker2021_Grid}. Based on the findings of that reference, we will use $r_b = 2000 r_{\mathrm{vdW}}$ in all our numerical calculations for \textsuperscript{7}Li.

Upon obtaining $u_{lc\bar{C}}^{\nu}(r)$ and $\varepsilon_{lc}^{\nu}$ we calculate the matrix elements $\matrixel*{k'l; \bar{C}'}{\hat{t}_{\alpha}^{l c}(z)}{kl; \bar{C}}$ directly from Eq.~\eqref{eq:LSbasis}. The momentum grid $k$ is composed using $N_k$ zeros of the spherical Bessel function, such that $j_l(k r_b) = 0$, up to a maximum two-body momentum $k_{\mathrm{max}}$. Once the transition matrix is obtained, we diagonalize to obtain the spectral expansion,
\begin{align}
\begin{split}
\matrixel*{k'l ; \bar{C}'}{\hat{t}_{\alpha}^{l c}(z)}{kl; \bar{C}} = \sum_{n} \braket*{k';\bar{C}'}{\chi_{l c}^{n}(z)} \tau_{lc}^{n}(z) \braket*{\chi_{l c}^{n}(z)}{k;\bar{C}},
\end{split}
\label{eq:tmatExp}
\end{align}
with eigenvalues $\tau_{lc}^{n}(z)$ and form factors $\braket*{k;\bar{C}}{\chi_{l c}^{n}(z)}$. Once more we suppress the Jacobi index $\alpha$ on the right-hand side. Note that we represent eigenfunctions of the two-body Hamiltonian with index $\nu$ and eigenfunctions of the transition matrix with index $n$, and will maintain this convention going forward.

The number of terms $n$ for each $l$, $c$ and a given value of the energy $z$ equals the number of grid points in $k$ times the number of two-body channels. As this number is typically too large to store the full transition matrix in memory, we sort the spectral expansion \eqref{eq:tmatExp} with respect to decreasing magnitude of $\tau_{lc}^{n}(z)$, and subsequently truncate the expansion at a term $n_{\mathrm{cut}}$, which sets the precision to which we include the transition matrix. In that case, Eq. \eqref{eq:tmatExp} represents a separable approximation to the formal transition matrix, which will allow us to write the three-body equations in the next sections into a one-dimensional form that can be efficiently solved on a computer \cite{Sitenko1991}. For further discussion of different separable approximations to the transition matrix and their consequences for studying Efimov physics, we refer the interested reader to Ref.~\cite{Mestrom2019}.

\subsection{Three-body recombination rate}
\label{sec:K3}

We are interested in the transition rate from a free zero-energy three-body state $\ket*{\Psi_{\mathrm{in}}}$ to an outgoing atom-dimer state $\ket*{(\vb{p}, c),  \varphi_{\nu}}$. The latter describes a state in which the pair is in the molecular state $\ket*{\varphi_{\nu}}$, while the third particle is in a plane wave state $\ket*{\vb{p}; c}$ with atom-dimer momentum $\vb{p}$ and spin state $c$. Such a process induces a particle density loss $\mathrm{d}n/\mathrm{d}t = - K_3 \ n^3$, where the recombination rate constant $K_3$ is determined as \cite{Lee2007,Moerdijk1996,Smirne2007},
\begin{align}
K_3 = \frac{12 \pi m}{\hbar} (2 \pi \hbar)^6  \lim_{E \rightarrow 0}  \sum_{{\nu},  c}   \int d\vu{p} \ p \abs*{\matrixel*{(\vb{p}, c),  \varphi_{\nu}}{\hat{U}_{\alpha0}(E)}{\Psi_{\mathrm{in}}}}^2.
\label{eq:K3}
\end{align}
Here $\hat{U}_{\alpha 0}(E)$ is the appropriate three-body transition operator. By inserting a complete set of three-body partial-wave basis states, one finds,
\begin{align}
\begin{split}
\int d\vu{p} \abs*{\matrixel*{(\vb{p}, c), \varphi_{\nu}}{\hat{U}_{\alpha0}(E)}{\Psi_{\mathrm{in}}}}^2 &= 4 \pi \sum_l \abs*{\matrixel*{\varphi_l^{\nu} pl; c}{\hat{U}_{\alpha0}(E)}{kp0; \bar{C}_{\mathrm{in}}  c_{\mathrm{in}}}}^2.
\end{split}
\end{align}
Here we have defined explicitly the $s$-wave incoming state $\ket*{\Psi_{\mathrm{in}}} = \lim_{E \rightarrow 0} \ket*{kp0;, \bar{C}_{\mathrm{in}} c_{\mathrm{in}}}$, and have decomposed the bound state $\ket*{\varphi_{\nu}}$ into partial-wave components $\ket*{\varphi_l^{\nu}}$ which capture the state of the pair. Accordingly, the full three-body state can then be written as $\ket*{\varphi_l^{\nu} p l; c}$.

To evaluate the matrix elements of $\hat{U}_{\alpha 0}(E)$, we solve the appropriate integral equations for three identical particles as introduced by Faddeev \cite{Faddeev1960},
\begin{align}
\hat{U}_{\alpha  0}(E) = \hat{G}_0^{-1}(E) + \hat{P} \hat{\mathcal{T}}_{\alpha }(E) \hat{G}_0(E) \hat{U}_{\alpha 0}(E),
\end{align}
known also as the Alt-Grassberger-Sandhas (AGS) equations \cite{Alt1967, Glockle1983}. Here $\hat{G}_0(E) = (E - \hat{H}_0)^{-1}$ is the uncoupled three-body Green's function, and $\hat{P} = \hat{P}_+ + \hat{P}_-$ is the sum of cyclic and anticyclic permutation operators of the three particle indices. The two-body problem of the previous section enters through the generalized two-body transition matrix, defined as,
\begin{align}
\hat{\mathcal{T}}_{\alpha }(E) = \hat{V}_{\alpha } + \hat{V}_{\alpha } \hat{G}_{\alpha }(E) \hat{V}_{\alpha },
\label{eq:Tmat3}
\end{align}
where $\hat{G}_{\alpha }(E) = (E - \hat{H}_0 - \hat{V}_{\alpha })^{-1}$. Note that by this definition $\hat{\mathcal{T}}_{\alpha }(E)$ is diagonal in the coordinates of the third particle, and its action on the pair is equivalent to the two-body transition matrix. The AGS equation can be rewritten to,
\begin{align}
\hat{U}_{\alpha  0}(E) = \frac{1}{3}\hat{G}_0^{-1}(E) (1 + \hat{P}) + \hat{P} \hat{\mathcal{T}}_{\alpha }(E) \hat{G}_0(E) \hat{U}_{\alpha  0}(E).
\end{align}
Since the first term vanishes on the energy shell, we may as well define a new operator $\hat{\bar{U}}_{\alpha  0}(E) = \hat{P} \hat{\mathcal{T}}_{\alpha }(E) \hat{G}_0(E) \hat{U}_{\alpha  0}(E)$, which then obeys,
\begin{align}
\hat{\bar{U}}_{\alpha  0}(E) = \frac{1}{3}\hat{P} \hat{\mathcal{T}}_{\alpha }(E) (1 + \hat{P}) + \hat{P} \hat{\mathcal{T}}_{\alpha }(E) \hat{G}_0(E) \hat{\bar{U}}_{\alpha  0}(E).
\end{align}
Finally, for identical particles it is valid to write $\hat{P} = 2 \hat{P}_+$, such that,
\begin{align}
\begin{split}
\hat{\bar{U}}_{\alpha  0}(E) = \frac{2}{3}\hat{P}_+ \hat{T}_{\alpha }(E) \left(1 + 2\hat{P}_+ \right) + 2 \hat{P}_+ \hat{\mathcal{T}}_{\alpha }(E) \hat{G}_0(E) \hat{\bar{U}}_{\alpha  0}(E).
\end{split}
\end{align}
This equation may now be used to evaluate the matrix element $\matrixel*{\varphi_l^{\nu} pl; c}{\hat{U}_{\alpha0}(E)}{kp0; \bar{C}_{\mathrm{in}}  c_{\mathrm{in}}}$, by inserting complete sets of three-body basis states $\ket*{kpl;\bar{C}c}$ between all operators. The required matrix elements of the transition operator read,
\begin{align}
\begin{split}
&\matrixel*{kpl ;\bar{C} c}{\hat{\mathcal{T}}_{\alpha}(E)}{k'p'l' ; \bar{C}' c'}  =  \frac{1}{4\pi} \frac{\delta(p-p')}{pp'} \delta_{ll'}   \delta_{c c'} \sum_n \braket*{k; \bar{C}}{\chi_{l}^n\left(Z_{c}(p) \right)} \tau_{l}^n(Z_{c}(p))\braket*{\chi_{l}^n\left(Z_{c}(p) \right)}{k'; \bar{C}'},
\end{split}
\end{align}
where we have inserted the spectral expansion in Eq.~\eqref{eq:tmatExp}. We have also defined,
\begin{align}
\begin{split}
Z_{c}(p) = E - \frac{3}{4} \frac{\hbar^2 p^2}{m} - \varepsilon_{c},
\end{split}
\end{align}
which represents the non-interacting energy of the pair. For brevity we suppress the third particle index $c$ for $\tau$ and $\chi$, as it can be inferred from the energy argument $Z_{c}$. To write the AGS equation we also need the matrix elements of $\hat{P}_+$, referred to as the recoupling coefficients \cite{Glockle1983}. We follow Ref.~\cite{Secker2021_2C} and decompose the operator as $\hat{P}_+ = \hat{P}_+^s \hat{P}_+^c$, where $\hat{P}_+^s$ acts only on the spin indices and $\hat{P}_+^c$ acts only on coordinates. The matrix elements of $\hat{P}_+^c$ are well known in literature and read \cite{Glockle1983},
\begin{align}
\begin{split}
 &\matrixel*{kpl}{\hat{P}_{+}^c}{k'p'l'} = \frac{1}{32 \pi^2} (-1)^{l'}  \sqrt{(2l+1)(2l'+1)}   \int_{-1}^{1}dx \frac{\delta\left(\sqrt{p'^2 + \frac{1}{4}p^2 + pp'x} - k \right)}{k \sqrt{p'^2 + \frac{1}{4}p^2 + pp'x} }  \frac{\delta\left(\sqrt{p^2 + \frac{1}{4}p'^2 + pp'x} - k' \right)}{k' \sqrt{p^2 + \frac{1}{4}p'^2 + pp'x} } \\ &   \hspace{8cm} \times P_l\left[\frac{\frac{1}{2}p^2 + p p' x}{p\sqrt{p'^2 + \frac{1}{4}p^2 + pp'x}} \right] P_{l'}\left[\frac{\frac{1}{2}p'^2 + p p' x}{p'\sqrt{p^2 + \frac{1}{4}p'^2 + pp'x}} \right],
\end{split}
\label{eq:Recouple3body}
\end{align}
where $P_l(x)$ is the Legendre polynomial. Upon inserting into the AGS equation and integrating away the $\delta$-functions, one finds that the recoupling of angular momenta can be conveniently expressed by the function,
\begin{align}
\begin{split}
\mathfrak{P}_{l l'}(p, q, x) = (-1)^{l'}  \sqrt{(2l+1)(2l'+1)}  P_l\left[\frac{\frac{1}{2}p^2 + p q x}{p \pi(q, p, x)} \right] P_{l'}\left[\frac{\frac{1}{2}q^2 + p q x}{q\pi(p,q, x)} \right],
\end{split}
\end{align}
where we have adopted the following notation from Ref.~\cite{Glockle1983},
\begin{align}
\begin{split}
\pi(p, q, x) = \sqrt{p^2 + \frac{1}{4} q^2 + pqx}.
\end{split}
\end{align}
The spin permutation operator may be calculated following the method of Ref.~\cite{Secker2021_2C}. For example, if we choose $\alpha = 3$ it is given as,
\begin{align}
\begin{split}
\matrixel*{\bar{C}_3 c_3;l}{\hat{P}_{+}^{\mathrm{s}}}{\bar{C}_3' c_3';l'}  &= \frac{1}{\sqrt{2(1 + \delta_{c_1 c_2})}} \frac{1}{\sqrt{2(1+ \delta_{c_1' c_2'})}} \bigg[ \delta_{c_1 c_2'} \delta_{c_2 c_3'} \delta_{c_3  c_1'} + (-1)^{l} \delta_{c_2 c_2'} \delta_{c_1 c_3'} \delta_{c_3  c_1'} \\ & \hspace{5cm} + (-1)^{l'} \delta_{c_1 c_1'} \delta_{c_2 c_3'} \delta_{c_3  c_2'} +  (-1)^{l + l'} \delta_{c_2 c_1'} \delta_{c_1 c_3'} \delta_{c_3  c_2'} \bigg]
\end{split}
\end{align}
Here it is useful to note that for single-channel interactions, where particles are always in identical spin states, $\hat{P}_+^s$ forbids coupling to dimer states with $l$ odd. Finally, to write the AGS equation as a linear set of algebraic equations we realize that the atom-dimer state can be written explicitly in terms of the eigenfunctions of the two-body transition matrix, following from the spectral expansion of Eq.~\eqref{eq:Tmat3},
\begin{align}
\begin{split}
\hat{\mathcal{T}}_{\alpha}(z) & \approx  \frac{\hat{V}_{\alpha}\ket*{\varphi_l^{\nu} p l; c}\bra*{\varphi_l^{\nu} pl; c}\hat{V}_{\alpha}}{z - \varepsilon_{\nu}^{l} - \frac{3}{4} \frac{\hbar^2 p^2}{m} - \varepsilon_{c}} \qquad \mathrm{for} \qquad z \approx \varepsilon_l^{\nu} + \frac{3}{4} \frac{\hbar^2 p^2}{m} + \varepsilon_{c}.
\end{split}
\label{eq:Tspec}
\end{align}
It follows that we can write,
\begin{align}
\begin{split}
\ket*{\varphi_l^{\nu} pl; c} = \hat{G}_0(E) \ket*{\chi_l^n\left(Z_{c}(p)\right)} \ket*{pl;c},
\end{split}
\label{eq:stob}
\end{align}
connecting each eigenstate $\nu$ of the two-body Hamiltonian with a unique term $n$ in the spectral expansion of the two-body transition matrix. 

We are now finally in a position to write the AGS equation in the partial-wave representation. For notational convenience we will write the recombination matrix element as,
\begin{align}
\begin{split}
\bar{U}_{lc}^{n}(p) = \matrixel*{\varphi_{l}^n pl; c}{\hat{\bar{U}}_{\alpha0}}{000;\bar{C}_{\mathrm{in}} c_{\mathrm{in}}},
\end{split}
\end{align}
which is stored numerically as a vector. The full AGS equation can then be written as,
\begin{align}
\begin{split}
\bar{U}_{lc}^{n}(p)&=  \frac{2}{3} \sum_{n'} \sum_{c'} \mathcal{Z}_{l0; c c'}^{n n'}\left(p,0 \right) \tau_{0}^{n'}\left(Z_{c'}(0) \right)\sum_{\bar{C}}\braket*{\chi_{0}^{n'}\left(Z_{c'}(0) \right)}{0;\bar{C}} \matrixel*{\bar{C}c'}{1+2\hat{P}_+^{\mathrm{s}}}{\bar{C}_{\mathrm{in}}c_{\mathrm{in}}} \\
 & + 8\pi \sum_{l'} \sum_{n'}  \sum_{c'} \int_0^{\infty} dq \ q^2 \ \mathcal{Z}_{ll'; c c'}^{n n'}\left(p,q \right) \tau_{l'}^{n'}\left(Z_{c'}(q) \right) \bar{U}_{c'}^{n'l'}(q).
\end{split}
\label{eq:AGSFull}
\end{align}
Here we have defined the kernel function as,
\begin{align}
\begin{split}
&\mathcal{Z}_{ll'; c c'}^{n n'}\left(p,q \right) = \frac{1}{2}\sum_{\bar{C} \bar{C}'} \int_{-1}^{1} dx \ \frac{\braket*{\chi_{l}^n\left(Z_{c}(p) \right)} {\pi(q,p,x); \bar{C}}\braket*{\pi(p,q,x); \bar{C}'}{\chi_{l'}^{n'}\left(Z_{c'}(q) \right)}}{E - \frac{\hbar^2}{m} \left(q^2 + p^2 + qpx \right) - E_{\bar{C}, c}} \mathfrak{P}_{l l'}(p,q, x) \matrixel*{\bar{C}c}{P_+^{\mathrm{s}}}{\bar{C}'c'}.
\end{split}
\label{eq:Zkern}
\end{align}
In these equations $q$ is an integrated atom-dimer momentum, and $x = \vu{p} \cdot \vu{q}$.

\subsection{Three-body bound states}

The energy of three-body bound states can be obtained from a similar but homogeneous integral equation \cite{Faddeev1960, Glockle1983, Sitenko1991}. It reads,
\begin{align}
\begin{split}
\ket*{\Phi_{\alpha}(E)} = 2 \hat{G}_0(E) \hat{\mathcal{T}}_{\alpha}(E) \hat{P}_+ \ket*{\Phi_{\alpha}(E)},
\end{split}
\end{align}
for a bound state $\ket*{\Phi(E)} = (1 + \hat{P}) \ket*{\Phi_{\alpha}(E)}$ at three-body energy $E<0$. Just like for the AGS/recombination equation, the bound-state equation can be expressed in the partial-wave three-body basis. For a separable transition matrix it can then be reduced to a one-dimensional integral equation by splitting off the dimer degrees of freedom as,
\begin{align}
\begin{split}
\braket*{kpl;\bar{C}c}{\Phi_{\alpha}(E)} = \sum_n \frac{\tau_l^n\left(Z_{c}(p)\right) \braket*{k;\bar{C}}{\chi_l^n(Z_{c}(p; E))} \Phi_{c}^{nl}(p; E)}{E - \frac{\hbar^2 k^2}{m} - \frac{3}{4} \frac{\hbar^2 p^2}{m} - E_{\bar{C}, c}}.
\end{split}
\end{align}
The bound-state equation becomes,
\begin{align}
\begin{split}
&\Phi_{lc}^{n}(p; E) =  8\pi  \sum_{l'} \sum_{c'} \sum_{n'}  \int_0^{\infty}  dq \ q^2 \mathcal{Z}_{ll'; c c'}^{n n'}\left(p,q \right) \tau_{l'}^{n'}\left(Z_{c'}(q)\right)   \Phi_{l'c'}^{n'}(q; E),
\end{split}
\label{eq:FaddeevFull}
\end{align}
again dependent on the kernel function defined in Eq.~\eqref{eq:Zkern}.

\subsection{Poles in the transition matrix}

As follows from Eq.~\eqref{eq:Tspec}, the function $\tau_l^n\left(Z_{c}(q)\right)$ has a pole whenever $q = \bar{q}_{lc}^n$, where,
\begin{align}
\begin{split}
\bar{q}_{l c}^n = \sqrt{\frac{4m}{3\hbar^2} \left(E - \varepsilon_{c} - \varepsilon_{l c}^n \right)},
\end{split}
\end{align}
correspondent to the momenta where the energy of the dimer subsystem is exactly equal to a two-body bound state. As is typical in scattering theory we circumvent these poles by adding an infinitesimal imaginary part $i0$ to the energy \cite{Taylor2006, Mestrom2019}, after which we can write,
\begin{align}
\begin{split}
\tau_{l}^{n}\left(Z_{c}(q \approx \bar{q}_{l c}^n) \right) \approx \frac{4m}{3\hbar^2} \frac{1}{(\bar{q}_{lc}^{n})^2 - q^2 + i0}.
\end{split}
\end{align}
This ensures that the integrals in the recombination and bound-state equations converge to a finite but complex value. In numerical practice we adjust our integral equations by subtracting and adding all integrals over the divergent parts of the transition matrix. The subtracted term then ensures convergence of the discretized integral, while the added term is evaluated analytically via the Sokhotski-Plemelj theorem, such that one obtains,
\begin{align}
\begin{split}
&8\pi \sum_{l'} \sum_{n'}  \sum_{c'} \int dq \ q^2 \ \mathcal{Z}_{ll'; c c'}^{n n'} \left(p,q \right) \tau_{l'}^{n'}\left(Z_{c'}(q) \right) \bar{U}_{l' c'}^{n}(q) =  \\ & \hspace{3cm} 8\pi \sum_{l'} \sum_{n'}  \sum_{c'} \int dq \ q^2 \ \mathcal{Z}_{ll'; c c'}^{n n'}\left(p,q \right) \tau_{l'}^{n'}\left(Z_{c'}(q) \right) \bar{U}_{l' c'}^{n'}(q) \\ 
& \hspace{3cm} - 8\pi \frac{4m}{3\hbar^2}  \sum_{l'} \sum_{n'}^{\mathrm{poles}}  \sum_{c'} \int dq (\bar{q}_{l' c'}^{n'})^2 \ \mathcal{Z}_{ll'; c c'}^{n n'}\left(p,\bar{q}_{l' c'}^{n'} \right) \frac{1}{\bar{q}_{l' c'}^{n'2} - q^2} \bar{U}_{l' c'}^{n'}(\bar{q}_{l' c'}^{n'}) \\
&  \hspace{3cm} + 8\pi \frac{4m}{6\hbar^2}  \sum_{l'} \sum_{n'}^{\mathrm{poles}}  \sum_{c'}  \bar{q}_{l' c'}^{n'} \ \mathcal{Z}_{ll'; c c'}^{n n'}\left(p,\bar{q}_{l' c'}^{n'} \right)    \left[\ln \left(\frac{q_{\mathrm{max}} + \bar{q}_{l' c'}^{n'}}{q_{\mathrm{max}} - \bar{q}_{l' c'}^{n'}} \right) - i \pi \right] \bar{U}_{l' c'}^{n'}(\bar{q}_{l' c'}^{n'})
\end{split}
\label{eq:PoleTerms}
\end{align}
Here the terms $n \in \mathrm{poles}$ are associated with dimer energies $\varepsilon_{l c}^n$ that fall in the range,
\begin{align}
\begin{split}
E - \varepsilon_{c} - \frac{3}{4} \frac{\hbar^2 q_{\mathrm{max}}^2}{m} < \varepsilon_{l c}^n < E - \varepsilon_{c} - \frac{3}{4} \frac{\hbar^2 q_{\mathrm{min}}^2}{m}. 
\end{split}
\end{align}
Here $q_{\mathrm{min(max)}}$ is the minimum(maximum) momenta in the numerical grid used for $q$ (see next section). At the momenta $\bar{q}_{l c}^n$ the three-body recombination process is exactly on-shell. Hence, to obtain the three-body recombination rate, we sum all $\abs{\bar{U}_{cl}^{n}(\bar{q}_{l c}^n)}^2$. For the bound state energies we use Eq.~\eqref{eq:PoleTerms} with $\bar{U}_{lc}^{n}(p) \rightarrow \Phi_{lc}^{n}(p)$.

\subsection{Numerical method and parameters}

To solve Eqs.~\eqref{eq:AGSFull} and \eqref{eq:FaddeevFull} we first discretize the atom-dimer momenta $p$ and $q$ using Gauss-Legendre quadrature \cite{Press2007}, segmented into three sections as $q_1 r_{\mathrm{vdW}} \in \left[10^{-5}, 10^{-2} \right]$, $q_2 r_{\mathrm{vdW}} \in \left[10^{-2}, 1 \right]$ and $q_3 r_{\mathrm{vdW}} \in \left[1, q_{\mathrm{max}} r_{\mathrm{vdW}} \right]$. Here $q_{\mathrm{max}}$ acts as an ultraviolet cut-off, which limits the maximum binding energy of dimers that couple to the three-body complex. After discretization, both equations may be written as straightforward matrix equations,
\begin{align}
\begin{split}
\vb{U} = \vb{Z}_0 +  \underline{Z} \ \underline{\tau} \ \vb{U},
\end{split}
\end{align}
where $\vb{U}$ is a vector containing the elements $\bar{U}_{l c}^{n}(p)$ or $ \Phi_{lc}^{n}(p)$. The vector $\vb{Z}_0$ contains the inhomogeneous term, which vanishes in the bound-state equation. The matrix $\underline{Z}$ contains the kernel function $\mathcal{Z}$, and the diagonal matrix $\underline{\tau}$ contains the values of $\tau$. In numerical practice, we are typically limited by the considerable memory required to store the matrix $\mathcal{Z}$, which restrains us both in the ultraviolet cut-off $q_{\mathrm{max}}$ and the maximum orbital quantum number $l_{\mathrm{max}}$. We can however reduce the dimension of the matrix equation by separating all terms into two distinct spaces, after which we can write a matrix equation for the terms in space 1 as,
\begin{align}
\begin{split}
\vb{U}_1 = \tilde{\vb{Z}}_0 +  \underline{\tilde{Z}} \ \underline{\tau}_1 \ \vb{U}_1,
\end{split}
\end{align}
where,
\begin{align}
\begin{split}
\tilde{\vb{Z}}_0 &= \vb{Z}_{0,1} + \underline{Z}_{12} \left(\underline{\tau}_2^{-1} - \underline{Z}_{22} \right)^{-1} \vb{Z}_{0,2} \\
\underline{\tilde{Z}} &= \underline{Z}_{11}  + \underline{Z}_{12}\left(\underline{\tau}_2^{-1} - \underline{Z}_{22}   \right)^{-1} \underline{Z}_{21},
\end{split}
\end{align}
Such a two-space approach is more generally used in quantum physics when one desires to eliminate a subspace of states whilst retaining the effects of their coupling to the states that remain, including for example the formal theory of scattering resonances \cite{Feshbach1962}, and the mapping between the Hubbard and Heisenberg Hamiltonians in condensed matter physics \cite{Mila2011}.

Assuming that the terms $\tau_2$ are small, we can make the first order approximation,
\begin{align}
\begin{split}
\tilde{\vb{Z}}_0 &\approx \vb{Z}_{0,1} + \underline{Z}_{12} \underline{\tau}_2 \vb{Z}_{0,2} \\
\underline{\tilde{Z}} &\approx \underline{Z}_{11}  + \underline{Z}_{12}\underline{\tau}_2 \underline{Z}_{21},
\end{split}
\end{align}
which significantly reduces the required memory as we do not need the elements $\underline{Z}_{22}$. Essentially the method above amounts to another reduction on the number of terms in the expansion \eqref{eq:tmatExp} for the transition matrix, although now the terms are included to first order instead of neglected entirely. 

\begin{figure}
\includegraphics{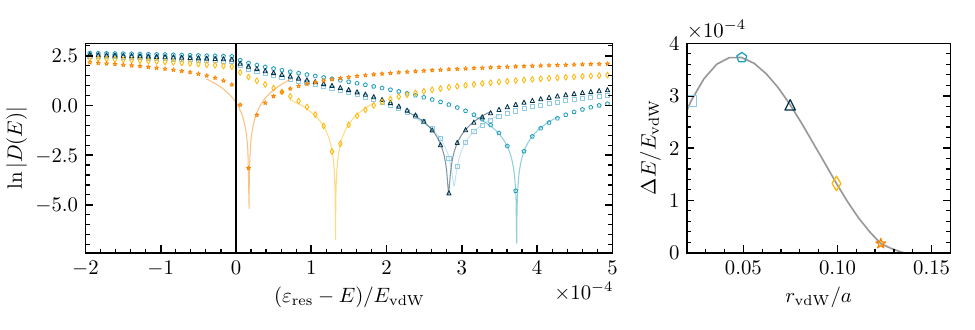}
\caption{\label{fig:DetPlot} Illustration of the numerical solution to the three-body bound-state equation for an Efimov resonance in a single-channel van der Waals potential near the 8'th two-body potential resonance. In the left-hand panel we plot scans of $\mathrm{ln}\abs{D(E)}$ as a function of the shifted three-body energy $\varepsilon_{\mathrm{res}} - E$, for a set of (positive) scattering lengths. Local minima in $\mathrm{ln}\abs{D(E)}$ signify the location of Efimov resonances, as follows from Eq.~\eqref{eq:DetFit}. Solid lines show fits of this equation to the minima. In the right-hand panel we show the resulting dependence of the trimer-dimer energy difference $\Delta E$ on the scattering length, where the scatter points match with the fits in the left-hand panel.}
\end{figure}

Although the bound-state equation looks superficially simpler than the recombination equation, finding its solution requires some iterative procedure in the three-body energy $E$ until one obtains a bound state energy. This iteration is complicated by the fact that the Efimov trimers we look for are embedded in the atom-dimer continua associated with deep dimer states. Hence, the Efimov state is not a true bound state but rather a three-body resonance with complex energy $E_* = \mathrm{Re}(E_*) - i \Gamma/2$, where $\Gamma$ is the energetic width of the state which defines a characteristic lifetime $\tau_{\mathrm{trimer}} = \hbar/\Gamma$. To simplify the search for bound states, we assume the Efimov state to be long lived enough such that $\mathrm{Re}(E_*) \gg \Gamma$. Then we scan the absolute logarithm of the determinant $D(E) = \mathrm{det} \left( \underline{\tilde{Z}} \underline{\tau}_1 - \mathbb{I}\right)$ along the real axis of $E$, noting that $D(E_*) = 0$. Upon passing this root, we can expand the determinant to lowest order in $\Gamma$,
\begin{align}
\begin{split}
\mathrm{ln}\abs{D(E)} \approx \mathrm{ln}  \abs{\dv{D}{E}}_{E = E_*} + \mathrm{ln} \left(\sqrt{(E - \mathrm{Re}(E_*))^2 + \frac{1}{4}\Gamma^2} \right).
\end{split}
\label{eq:DetFit}
\end{align}
It follows that the location of a minimum in $\mathrm{ln}\abs{D(E)}$ corresponds with the value of $\mathrm{Re}(E_*)$, which we extract to obtain the energy of the Efimov trimer. Note that this approach will only work if the root in the determinant lies close enough to the real axis such that Eq.~\eqref{eq:DetFit} is valid. Following experiment \cite{Yudkin2024}, we will be primarily interested in the trimer-dimer energy difference $\Delta E = \varepsilon_{\mathrm{res}} - \mathrm{Re}(E_*)$, measuring the three-body energy relative to the energy of the Feshbach dimer $\varepsilon_{\mathrm{res}}$. To illustrate our method and the general behavior of $\Delta E$ near a two-body resonance, we plot in Fig.~\ref{fig:DetPlot} scans of $\mathrm{ln}\abs{D(E)}$ along the appropriately shifted energy axis $\varepsilon_{\mathrm{res}} - E$. We use here a single-channel van der Waals potential at the 8'th $s$-wave potential resonance \cite{Kraats2023}, allowing us to compute the determinant at a large number of points thus illustrating the appearance of sharp local minima which we identify as Efimov resonances, fitted well by Eq.~\eqref{eq:DetFit}. Plotting the resulting values of the trimer-dimer energy difference $\Delta E$ as a function of scattering length then shows the characteristic behavior of the Efimov trimer, merging with the atom-dimer continuum at the critical scattering length $a_* \approx 1.1 a_+$ \cite{Braaten2006, Gross2011}.

\subsection{Spin models}

An inherent advantage of our method, based on integral equations, is that it is reasonably simple to exclude specific three-body channels from the computation if so desired. Up until this point, the method that we have formulated is quantum mechanically rigorous, except for the omission of non-additive three-body forces, which we assume to be negligible in the systems of interest \cite{Naidon2017}. We will refer to this as the Full Multichannel Spin (FMS) model \cite{Secker2021_39K, Li2022}. In the literature, FMS calculations are rare, and it is instead common practice to reduce the spin degrees-of-freedom of the total system by various approximations, see e.g. \cite{Petrov2004, Gogolin2008, Massignan2008, Schmidt2012,  Wang2014, Wolf2017, Langmack2018, Chapurin2019, Secker2021_2C, Haze2022}. We will study a specific simplifying approach to the spin structure known as the Fixed Spectating Spin (FSS) approximation \cite{Secker2021_39K}, which turns off the two-body interaction whenever the third particle (the spectator) is not in the incoming spin state $c_{\mathrm{in}}$. Formally, this is achieved through the replacement $\hat{V}_{\alpha} \rightarrow \hat{V}_{\alpha}^{\mathrm{FSS}}$, where,
\begin{align}
\begin{split}
\hat{V}_{\alpha}^{\mathrm{FSS}} = \hat{V}_{\alpha} \dyad*{c_{\mathrm{in}}}{c_{\mathrm{in}}}.
\end{split}
\end{align}
It follows that the two-body transition matrix similarly vanishes whenever the spectating particle has changed spin, meaning that in both the recombination and bound-state equations we can simply set $c \rightarrow c_{\mathrm{in}}$ everywhere. This leads to a significant reduction in the dimension of the integral equations, such that the resulting numerical computation is much less expensive. The validity of the FSS approximation relies on the characteristic suppression of three-body probability in the short-range, which should make it relatively unlikely for three particles to approach within the spin-exchange regime simultaneously. Numerical studies have shown however that for some systems, the effects of three-body spin-exchange where all three particles change their spin can still be significant \cite{Secker2021_39K, Li2022}, and are essential in reproducing the correct three-body parameter for \textsuperscript{7}Li specifically \cite{Kraats2024}. In the next section we will expand on these results, comparing the FSS and FMS approaches also for positive scattering lengths.

\section{Results}
\label{sec:results}

As an application of our model, we will study the rate of three-body recombination $K_3$ near two high-field Feshbach resonances in \textsuperscript{7}Li, whose characteristic properties we collect in table \ref{tab:fesres}. Both resonances are of intermediate width with  $s_{\mathrm{res}} \sim 1$ \cite{Chin2010}, well into the non-universal regime for the three-body parameter \cite{Kraats2023}. 
\begin{table}[]
\begin{tabular}{cccccc}
\hline
\hline
State      & $B_0$ {[}G{]} & $a_{\mathrm{bg}}$ $[a_0]$ & $\Delta B$ {[}G{]} & $s_{\mathrm{res}}$ & $M_F$ \\ \hline
$aa$ & 738.2(2)      & -18.24                & -237.8             & 0.56              & $3$ \\
$bb$ & 893.7(3)      & -20.98                & -171.0             & 0.493             &  $0$ \\
\hline
\hline
\end{tabular}
\caption{Parameters of the two Feshbach resonances studied in this work \cite{Gross2011, Dyke2013, Jachymski2013}. Here $B_0$ is the resonant magnetic field, $a_{\mathrm{bg}}$ the background scattering length, $\Delta B$ the magnetic field width, $s_{\mathrm{res}}$ the resonance strength parameter\cite{Chin2010}, and $M_F$ the total magnetic projection quantum number for the three-body calculation.}
\label{tab:fesres}
\end{table}
As we are interested in Efimovian signatures, it is useful to recast our results for $K_3$ using the theoretical prediction $K_3(a) = 3 C_{\pm}(a) \hbar a^4/m$ \cite{Braaten2006}. Here the function $C_{\pm}(a)$ captures the Efimovian modulation of the universal $a^4$ scaling of $K_3$ near resonance, for positive ($+$) and negative ($-$) scattering lengths respectively. Closed form expressions for $C_{\pm}(a)$ can be obtained within a zero-range effective field theory upon introducing a three-body parameter \cite{Braaten2006, Kraemer2006, Gross2011},
\begin{align}
C_{-}(a) &= 4590 \frac{\mathrm{\sinh}(2 \eta_-)}{\mathrm{sin}^2(s_0 \ln(a/a_-)) + \mathrm{sinh}^2(\eta_-)},
\label{eq:Cm}
\end{align}
\begin{align}
C_{+}(a) &= 67.1 e^{-2 \eta_+} \left(\mathrm{cos}^2(s_0 \ln(a/a_+)) + \mathrm{sinh}^2(\eta_+) \right) + 16.8(1 - e^{-4 \eta_+}).
\label{eq:Cp}
\end{align}

\begin{figure}
\includegraphics{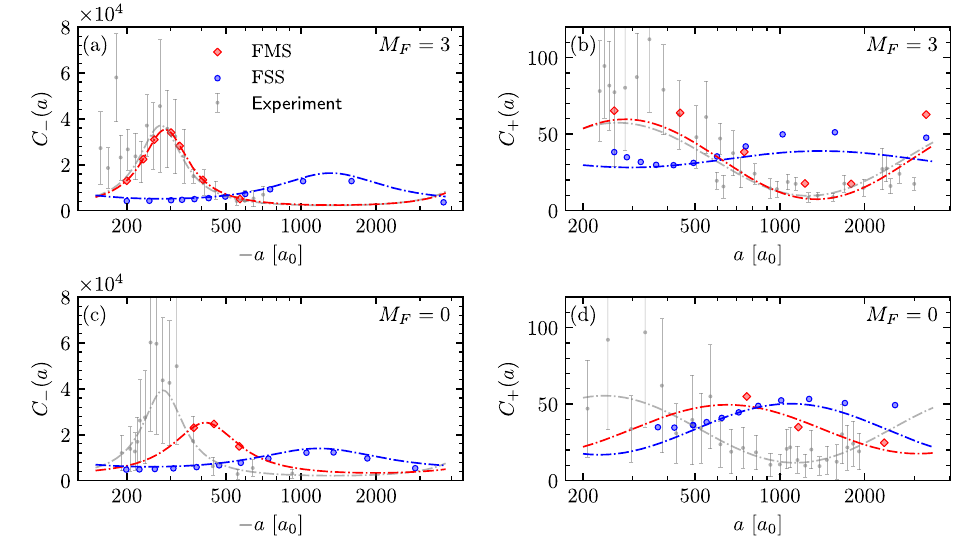}
\caption{\label{fig:C_negpos} Efimovian enhancement $C_{\pm}(a)$ of the three-body recombination rate near both Feshbach resonances in table~\ref{tab:fesres}. FMS results (red) and FSS results (blue) are compared with the experimental data of Ref.~\cite{Gross2011} (grey). Dash-dotted lines show fits of the data to the predictions of effective field theory in Eqs.~\eqref{eq:Cm} and \eqref{eq:Cp}, which we use to extract the three-body parameters $a_{\pm}$ and trimer widths $\eta_{\pm}$.}
\end{figure}

We will assume a priori that our calculated values of $K_3(a)$ obey the $a^4$ power law, and subsequently extract the associated Efimovian functions $C_{\pm}(a) = mK_3(a)/3\hbar a^4$. We can then fit the data directly using Eqs.~\eqref{eq:Cm} and \eqref{eq:Cp}, thus checking the validity of these universal expressions in \textsuperscript{7}Li, and obtaining explicit values for the three-body parameters $a_{\pm}$, with associated trimer widths $\eta_{\pm}$. The results, for both Feshbach resonances, are shown in Fig.~\ref{fig:C_negpos} and table \ref{tab:3BP}. Here we note that the results for $K_3$ for negative scattering lengths have been published previously elsewhere \cite{Kraats2024}.

\begin{table}[]
\begin{tabular}{ccccccc}
\hline
\hline
                           State & Model & $a_-/r_{\mathrm{vdW}}$ & $\eta_-$ & $a_+/r_{\mathrm{vdW}}$ & $\eta_+$ & $a_+/\abs{a_-}$  \\ 
                           \hline
&FSS  & -40.5                  & 0.63     & 43.1                   & 0.91   &  1.06                 \\
$M_{F} = 3$&FMS & -8.79                  & 0.27     & 8.69                  & 0.13 & 0.99                      \\
&Experiment\cite{Gross2011}  & -8.43(37) & 0.253(62) & 8.16(49) & 0.170(41) & 0.97(8) \\
\hline
&FSS  & -36.3                  & 0.78     & 33.6                   & 0.35 &   0.92                    \\
$M_{F} = 0$&FMS  & -12.7                  & 0.38     & 20.0                   & 0.37  & 1.57         \\
&Experiment\cite{Gross2011}   & -8.62(37) & 0.180(48) & 7.33(77) & 0.213(79) & 0.85(11) \\
\hline
\hline
\end{tabular}
\caption{Three-body parameters $a_{\pm}$ and trimer widths $\eta_{\pm}$ obtained from the fits to universal expressions for $C_{\pm}(a)$, as shown in Fig.~\ref{fig:C_negpos}, compared with similar fits to the experimental data.}
\label{tab:3BP}
\end{table}
Our results show significant differences between FSS and FMS spin models, clearly indicating that the FSS approximation is inadequate for \textsuperscript{7}Li atoms in the strongly interacting regime. For negative scattering lengths, where no two-body bound state exists, we find that the FSS approximation produces a broad Efimov resonance at relatively large scattering length, in accordance with the general expectation that the Efimov trimer is shifted towards the three-body threshold for intermediate strength Feshbach resonances \cite{Schmidt2012, Langmack2018, Secker2021_2C, Kraats2023}. In contrast, upon going to the FMS model we observe a dramatic shift of the position of the Efimov resonance to lower absolute scattering lengths, with the three-body parameter for the $M_{F}=3$ case even receeding \textit{below} the universal van der Waals value. The resonance itself is also significantly narrowed, indicating a non-trivial stabilizing effect of the additional spin channels on the Efimov trimer. Both these effects significantly improve on the match between the theoretical calculations and the experimental data, indicating the strong presence of three-body spin-exchange in experiments, and the accuracy of our numerical model. One will note however that for the case $M_{F} = 0$, a significant discrepancy between the experimental data and FMS results remains. We attribute this mismatch to a lack of convergence of our numerical method, originating from the fact that the $M_{\mathrm{tot}} = 0$ channel has a much larger number of coupled three-body channels, which limits us in the choice of numerical parameters. Specifically, for the $M_{F} = 3$ channel we use $l_{\mathrm{max}} = 12$ and $q_{\mathrm{max}}r_{\mathrm{vdW}} = 40$, which we have confirmed gives adequate convergence in the value of $a_-$. For the $M_{F} = 0$ channel numerical limitations force us to reduce the maximum partial wave to $l_{\mathrm{max}} = 8$, which is not sufficient to fully converge the result, thus explaining the mismatch with the experiment. For further discussion of this point we refer the reader to the Supplemental Material of Ref.~\cite{Kraats2024}. 

For positive scattering lengths, we find once more that the value of $a_+$ is decreased significantly in the FMS model compared to an FSS calculation. For the $M_{F} = 3$ channel, this shift again leads to an excellent match with the experiment of Ref.~\cite{Gross2011}, while the associated FSS results fail to reproduce the data to any satisfactory degree. In addition, the ratio $a_+/\abs{a_-}$ of our FMS results matches well with both the experimental data and the theoretical prediction $a_+/\abs{a_-} = 0.96(3)$ \cite{Braaten2006, Gross2011}. A similar shift of the Efimov resonance is seen in the $M_F = 0$ channel, although as before the numerical complexity in this more highly coupled three-body channel does not allow us to fully converge the calculations. As for negative scattering lengths however, the general trend of our numerics suggest that a converged calculation will give a further decrease of $a_+$, towards the experimental data.

Next to the comparison with experiment, our results for positive scattering length additionally reveal that the universal prediction for $C_+(a)$, as given in Eq.~\eqref{eq:Cp}, can not give a good fit to the recombination data over the full range of scattering lengths we study, indicating the presence of finite range effects which are unaccounted for in the universal theory. This effect is especially prevalent for the FSS results, where the fit of Eq.~\eqref{eq:Cp} is very poor leading also to a large uncertainty in $a_+$ and $\eta_+$. This observation however, is fully in line with the experimental data, which is also quite poorly fitted by the universal expression, and actually finds better match with our numerical computations. Similarly, in Ref. \cite{Dyke2013} it was found that a good fit of the Efimov resonance for $a>0$ could only be obtained by adding an overall scaling factor to Eq.~\eqref{eq:Cp}, which is also suggested by our FMS data in the $M_F = 3$ channel. These discrepancies indicate that an extended form of Eq.~\eqref{eq:Cp} which accounts for finite range corrections may be required to match the data. Approaches for adding such corrections to the universal Efimovian theory have indeed been discussed in the literature, see for example Ref.~\cite{Gattobigio2019}.

\section{Conclusion}
\label{sec:conc}

In this article we have described a state-of-the-art multichannel method for computing the three-body recombination rate and three-body bound state energy in a system of alkali-metal atoms near a Feshbach resonance, using realistic molecular pairwise interaction potentials. We have subsequently applied this method to study Efimovian signatures near two experimentally relevant Feshbach resonances in \textsuperscript{7}Li, where we show strong influence of spin-exchange effects which are typically neglected in the three-body literature. In particular, we find that such effects are essential in reproducing experimental data.

Our work poses several opportunities and challenges for further research. First it is clear that the numerical complexity of simulating interacting three-body systems at the level of accuracy required to reproduce experiment is considerable, requiring expensive high-performance computing facilities and considerable time investments. These points motivate further searches for accurate simplifying assumptions or improvements on the numerical efficiency of our method, which may then allow us to fully converge all numerical computations and thus close the remaining gaps between theory and experiment, which are highlighted once more in the present article. It may also allow for an application of the bound state method introduced in this work to realistic multichannel systems, which poses an even more challenging problem numerically due to the larger number of three-body energies required. 

Looking towards the future, our model can be applied generally in the search for non-universal features in three-body physics in ultracold atomic gases. We have shown that our method, once converged, accurately reproduces the experimentally observed three-body recombination rate, even when this rate deviates strongly from the universal expectation. We have thus developed a tool which can be used to guide future experiments, by predicting the three-body loss rate in general systems to a high degree of accuracy. For example, quantum gas experiments and technologies are typically limited by the lifetime of the gas \cite{Kunimi2019, Ratzel2021}, and thus may wish to exploit non-universal minima in the three-body loss rate, which can be searched for using our method. Finally, it is interesting to extend our method to also capture elastic three-body scattering, which is known to significantly affect the phase diagram of Bose-Einstein condensates in the regime of small scattering lengths, and can provide a mechanism for the stabilization of quantum droplets \cite{Bulgac2002, Tan2008, Mestrom2019_2, Mestrom2020}. The direct observation and characterization of this scattering remains an elusive challenge to contemporary experiments, making accurate numerical models invaluable.

\section*{Acknowledgments}

This work is financially supported by the Dutch Ministry of Economic Affairs and Climate Policy (EZK), as part of the Quantum Delta NL program, and by the Netherlands Organisation for Scientific research (NWO) under Grant No. 680.92.18.05. The results presented in this work were obtained on the Dutch national supercomputer Snellius, with support from the Eindhoven Supercomputing Center and SURF.

\section*{Statements and Declarations}

The authors have no competing interests to declare that are relevant to the content of this article.

\bibliography{References.bib}

\end{document}